\documentclass[12pt,journal,compsoc]{IEEEtran}

\usepackage{url}

\usepackage{graphicx}
\ifCLASSINFOpdf
  \usepackage{graphicx}
\else

\fi

\usepackage[official]{eurosym}

\begin{document}

\title{Anonymous online purchases with exhaustive operational security}

\author{Vincent Van Mieghem,
        Johan Pouwelse}

\IEEEcompsoctitleabstractindextext{%
\begin{abstract}
This paper describes the process of remaining anonymous online and its concurrent operational security that has to be performed. It focusses particularly on remaining anonymous while purchasing online goods, resulting in anonymously bought items. Different aspects of the operational security process as well as anonymously funding with cryptocurrencies are described. Eventually it is shown how to anonymously purchase items and services from the hidden web, as well as the delivery. It is shown that, while becoming increasingly difficult, it is still possible to make anonymous purchases. Our presented work combines existing best-practices and deliberately avoids untested novel approaches when possible.
\end{abstract}

}

\maketitle

\IEEEdisplaynotcompsoctitleabstractindextext

\IEEEpeerreviewmaketitle

\section*{Introduction}
\begin{figure*}[!htb]
  \includegraphics[width=\textwidth]{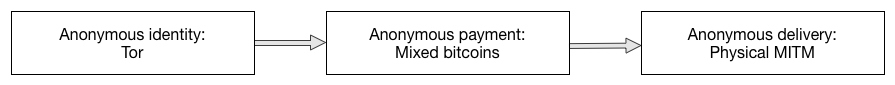}
  \caption{Elements of anonymous purchasing process.}
\end{figure*}

\IEEEPARstart{W}{e} present a detailed approach to anonymously purchase goods and services online. 

As human life shifts towards the world wide web, the interest of large Internet companies and government agencies in the Internet increases significantly. Remaining anonymous on the Internet has ever before been as hard as it is today. Every webpage visited, every file downloaded, every email or message sent, every purchase made, can be traced back and read by other parties. This paper describes the methods of remaining anonymous on the Internet and the concurrently required operations security processes. The first section explains the basic principles and tools regarding the operations security, also known as OPSEC. Thereafter, the process of obtaining anonymous bitcoins is described and finally the methods of purchasing online goods is explained.

\section{Extensive operational security}
Remaining anonymous on the internet is a though process. Anonymity cannot be obtained through just the usage of certain tools or services. It is a process that comes by keeping every single  asset involved in the anonymous operation, anonymous and minimise the amount of incriminating evidence on those assets. Operational security, hereafter OPSEC, plays a critical role in remaining anonymous online.

Anonymity does not work retroactively. At the beginning of an anonymous operation, the anonymity of all the assets involved have to be thought of very carefully. The environment that will protect the anonymity has to be set up thoroughly, from the beginning. The following subsections go into depth on the principles that ensure a high level of anonymity.

\subsection{Compartmentation}
One of the key principles of OPSEC is applying compartmentation. Compartmentation is the limiting of the ability for persons or entities to access information needed to perform certain tasks and isolating information as much as possible. In this case it means that all the digital evidence and information should be (if possible physically) separated from each other so that linking pieces of information is hard and cannot be used to extend the image of the anonymous identity that a person is using. Straight forward examples are:

\begin{itemize}
	\item Possessing and using different online accounts (i.e. Facebook or IRC nicknames) to sign up for anonymous services online.
	\item Using different anonymous (prepaid) credit card accounts to make purchases. 
	\item Removing XMP metadata (that may contain geolocation data) from photo's that a person uploads during the operation. LulzSec member W0rmer was exposed by not removing XMP data from a picture that contained geo-location of his girlfriend \cite{w0rmer}, which linked to his real identity. 
\end{itemize} 

But compartmentation goes as far as:
\begin{itemize}
	\item Using numerous hardware, i.e. mobile phones, laptops and CD-R's (USB-devices contain microcontrollers that can be reprogrammed into a spying device\cite{badusb}), for doing different operations or components of an operation.
	\item Spoofing MAC-addresses that are used to identify hardware.\cite{hammond}
	\item Using different physical internet connections for different parts of an operation, i.e. bars, hotels, open or hacked WiFi networks. LulzSec hacker Jeremy Hammond was caught by identifying his laptop using the MAC-address found on his personal WiFi network\cite{hammond}.
	\item Doing parts of an operation in different timezones, including never discussing environmental aspects, i.e. weather conditions. Hammond also revealed a large amount of profiling information during IRC chats, that were used to personally identify him.\cite{hammond}
	\item Removing any type of logs, i.e. syslog, chat logs from the operating system.
\end{itemize}
It should not be possible for an observer of a persons online activity to link properties of different compartments to each other. Every step or move online has to be advised of not linking compartments and extending the online image of the real identity. This effect is referred to as contamination. To prevent this, the general principle `proactively paranoid' is often used in OPSEC. 

\subsection{Online identity}
Depending on the kind of operation, the fake identity that will be used, has to be as authentic as possible. A layered approach is used, meaning that one would create a fake online identity and completely compartmentise this identity from its real identity. This fake identity would then be used to create other fake identities. It ensures that if one fake identify gets compromised, it would not lead to de-anonymization of the person's real identity, but instead just one `layer' or `compartment' of the identity protection would have been `peeled off'. In practice this means that created email addresses point consequently only to the email address of its previous `layer' and not layers beneath its previous `layer'. 

As in other OPSEC practices, avoiding contamination and profiling between the `wrapped' identities is vital.     

\subsection{Linking assets in real life}
The purchase of online goods will in almost every case, involve the physical contact of the person with entities that may be able to generate identifying material of the person's real identity. In general one should always try, for every asset involved, to prevent purchases in stores with camera surveillance and, obviously, never use electronic payment methods that may link to the real identity.

\subsubsection{Purchase of hardware}
In general, purchasing any kind of hardware via online marketplaces\footnote{http://marktplaats.nl}\footnote{https://ebay.com} is good practice (unless the hardware is already blacklisted due to past usage in compromised operations). Second hand hardware has reduced tracking perspectives which otherwise may lead to the de-anonymisation of the person's real identity. 
By purchasing hardware in stores, the person becomes vulnerable to in-store camera systems. To minimise exposure, only purchase from small local stores and start to use the hardware only a few months after. If unique hardware ID's may link to the time and location of purchase, the probability that the owner of the store has already removed video evidence of the moment of purchase, is larger. Making purchases in extreme weather conditions and dressing accordingly, may provide extra cover without drawing attention, i.e. sunglasses/scarf and cap. 
If hardware is bought online, it must never be delivered to the person's address or be paid with electronic payment systems other than anonymous prepaid credit cards and anonymised cryptocurrencies. It should be noted that spendings of prepaid credit cards are monitored\cite{the_grugq}. 

\subsubsection{Mobile Phones and SIM cards}
Only when usage of mobile phones in an operation is mandatory, they should be used. Due to easy localisation\cite{MONKEYCALENDAR} and decryption of communications\cite{SIM-hack}, this medium is considered insecure. Mobile phones should be disposable, featuring a minimal OS and functionalities, resulting in a reduced attack surface. The phones should be used in combination with multiple sim cards. Phones that will not be used anymore should be completely destroyed. While phones are not used, the devices should be battery unplugged or stored in a leaden box to prevent emission of signals, especially when a smartphone or WiFi device is used. The phones should never be used to make phone calls under the real identity of the person. Again, compartmentise and avoid contamination of those compartments. 

WiFi hotspots may keep logs of devices that are in range. Even unconnected WiFi devices may be logged at point of operation. This may link back to the persons real identity.

\subsection{Secure communication}
Secure communication is one of the most vital assets in an operation and if well understood, the easiest component to perform correctly in an anonymous operation. In general, open-source software projects that are end-to-end encrypted are used for secure communication. The methods described in the following sections are, as of the moment of writing, extremely secure, as concluded from N.S.A. documents from the Snowden-leaks\cite{Spiegel}.

\subsubsection{Email}
One or multiple anonymous email accounts are required that do not have personal linked information in them. Safe-mail\footnote{https://safe-mail.org} is a provider that provides email with minimal personal information upon sign-up. Gmail also allows for easy sign-up, although is more monitored\cite{Gmail-monitored} and sign-up over Tor requires phone verification.
This email account will be used together with GnuPG encryption of e-mails. Usage of keys longer than 4096-bit is important and authentication tokens to protect the GnuPG private key provide an extra layer of security. Descriptive subjects should be avoided, as subjects of e-mails are \emph{not} encrypted.

\subsubsection{IM}
For Instant Messaging, a protocol named Off-The-Record (OTR)\cite{OTR} can be used. This is a XMPP-based protocol, that can be used to communicate over Jabber services, i.e. securejabber.me or CCC.de XMPP service\footnote{jabber.ccc.de}. Password renewal on regular basis is important practice. XMPP services that provide self-signed certificates, signed with their GnuPG private key, provide an extra layer of security against PKI attacks\cite{CA-mitm}.

\subsubsection{Voice}
Open Whisper Systems is a group of open-source contributors that build end-to-end secure communication tools. RedPhone/Signal is a smartphone application uses the ZRTP protocol to provide end-to-end VoIP. Together with TexSecure, it provides perhaps the strongest security in mobile communications\cite{privacy-apps-snowden}. However, it should be noted that use of these techniques implicates vulnerability for profiling.\cite{skynet}

\subsection{Hardware \& Operating System}
The computer used to enter the internet, should be as lean and mean as possible so that possibility to store incriminating evidence is minimised. This means a laptop should be used with, preferably a pre-Intel 945\footnote{"Active Management Technology" is a technology built into Intel chipsets after the Intel 945 that allows for remote access.} no battery (easy shut down is necessary), no harddisk, WiFi chip, bluetooth, microphone and webcam. Try to always use an ethernet connection, that as extra fail-safe is routed over a Tor router like P.O.R.T.A.L.\footnote{https://github.com/grugq/portal}. This ensures that all Internet traffic is routed over the Tor network. The Tails operating system is designed to refrain the system of storing logs and any files and by default routes all the traffic automatically over Tor. Persistent encrypted storages with plausible deniability can be used to store data using TrueCrypt\footnote{http://istruecryptauditedyet.com}. 

Tor should always be used as it grants a good level of anonymity. The anonymity of a VPN is only as strong as the provider is willing to protect users data against law enforcement.\cite{CAK} VPN's provide privacy of the users data, but does not provide anonymity. Therefor, always connect to a VPN via Tor, if a VPN is necessary.

Another method of remaining even more anonymous is to route the traffic over a botnet, and using one of the infected nodes of the botnet as exit nodes\footnote{It should be carefully noted that in some countries, this method of traffic anonymization may be illegal.}.

\section{Funding wallets with anonymous bitcoin}
The currency that provides the most anonymity if used well, is Bitcoin. It should be carefully noted that all the transactions made are transparent in Bitcoin blockchain. Therefore, it is important to add a layer of anonymity. Green et al. have shown that it is possible to build this into Bitcoin itself and designed Zerocash\cite{zerocash}. This section explains how to practically accompany anonymity between the bitcoin wallet and the owner.

\subsection{Creation of anonymous wallets}
Creation of anonymous wallets can be done using multiple methods. Using bitaddress.org a user is able to generate a valid Bitcoin wallet. This website provides an in-browser method using 278 random user input points. However, this is by far the most insecure way of creating anonymous wallets. It has to be trusted that the JavaScript served has not been compromised, either server-side or in transit. Also, the browser itself should not be compromised when executing the JavaScript. 

More secure methods of wallet creation are using self built versions of open source Bitcoin wallets Armory\footnote{https://bitcoinarmory.com} or Electrum\footnote{https://electrum.org} on an offline machine that has never touched the internet directly. 
\begin{figure}[htb]
\includegraphics[height=32mm]{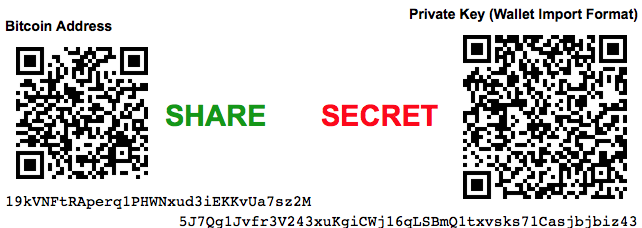}
\caption{Bitcoin wallet, consisting our of public and private component. Created using https://bitaddress.org}
\end{figure}

For an extra layer of security, cold storage wallets should be used. In this method, the private key of the wallet is air-gapped from the online machine that is used to connect to the blockchain.

\subsection{Physical anonymous meetup}
Two methods for funding wallets that involve physical exposure are explained. The first method is the most anonymous method, if performed correctly. It consist of a physical meetup, set up via localbitcoins.com. Buyer (using anonymous identity) and seller agree on a rendezvous point. The buyer pays the seller with cash and the seller transfers the agreed amount of bitcoins to an anonymous wallet on the spot. It is important to ensure that the rendezvous point is safe against robberies, but is not monitored by law enforcement, i.e. CCTV camera systems.

\subsection{Bank wire using fake identify}
Obtaining a fake identity allows for creation of anonymous bank accounts. These accounts can then be used to buy bitcoins through services like Coinbase\footnote{https://www.coinbase.com} and Circle\footnote{https://www.circle.com/en}. Obtaining a fake passport is enough to open a bank account that can then be used to fund bitcoin wallets. A fake passport can be bought through the hidden web, explained in section \ref{sec:purchasing}. 

\subsection{Bitcoin mixing}
Bitcoin `mixing', also known as `tumbling' or `laundering', is the process of anonymising bitcoins. This is achieved by randomly moving and shuffling small parts of bitcoins to a large amount of wallets, with the intention of confusing the trail back to the funds' original source. Third parties offer this anonymising service. The `Big Three' mixing services consists of BitBlender\footnote{bitblendervrfkzr.onion}, Bitcoin Fog\footnote{fogcore5n3ov3tui.onion} and Helix\footnote{grams7enufi7jmdl.onion/helix}. Other derivatives of the mixing protocols exist, i.e. SharedCoin\footnote{http://sharedcoin.com} and Bitmixer\footnote{bitmixer2whesjgj.onion} are examples of employing a slightly different protocol. Some mixing services may have a clear web entrance as well, but due to possible interception of traffic at Tor exit-nodes implying the risk of MITM, it is recommended to always use the hidden entrance of the service, as the connection is then peer-to-peer encrypted.
In this project, Bitmixer was used to anonymise bitcoins, in which users benefit from the instantly available mixed bitcoins. Bitmixer uses a large reserve of bitcoins to exchange a user's bitcoins with other bitcoins from the FIFO reserve. This reserve is assembled from a large amount of other transactions to Bitmixer. By using a time delay of delivery, multiple delivery wallets (also referred to as `forward addresses'), and a custom additional fee, in-out analysis of the Bitmixer reserve is prevented. 
In addition to that, Bitmixer uses a private `bitcode' that identifies a batch of previously owned bitcoins by a user. This ensures that a user will never receive its previously owned bitcoins after processing by Bitmixer. 

\begin{figure}[htb]
\includegraphics[width=90mm]{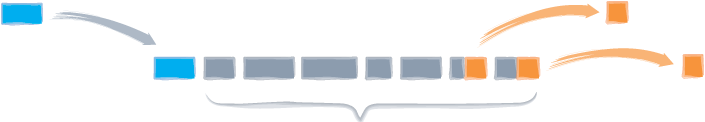}
\caption{Bitmixer stash, consisting a persons bitcoins (blue) and the anonymised bitcoins returned by Bitmixer.}
\end{figure}

\section{Purchasing Goods and Services}
\label{sec:purchasing}
The purchase of goods and services can be either via the clear web or the hidden web, as long as ensured that the visit of the websites is done using the previously described methods and that payment is done using multiple wallets of which the bitcoins themselves are again mixed to prevent analysis of purchase patterns.

The purchase of items on the hidden web often establishes via marketplaces on the hidden web, called `Dark Markets'. Popular Dark Markets are Agora, Middle Earth Marketplace, Outlaw Market and Evolution\footnote{Evolution was the most popular marketplace after Silk Road 1 \& 2 were closed by the FBI. Evolution shut down on March 18 2015 in an apparent `exit scam', in which the site's administrators shut down their market abruptly in order to steal the bitcoins kept in users' escrow accounts. The administrators stole Bitcoins for an estimated amount of \$15 million USD.}\cite{evolution}. In general, marketplaces function as an independent `trusted' Man in the Middle for a transaction between vendor and buyer. The buyer transfers the bitcoins to a wallet maintained by the site. The vendor is then `ensured' that the funds are available for purchase. Upon reception of of goods by the buyer, the site transfers the bitcoins to the vendor. Furthermore, Dark Markets are highly dependent on reputation of vendors. This guarantees a certain amount of trustworthiness of the vendor.
\begin{figure}[htb]
	\begin{center}
		\includegraphics[width=30mm]{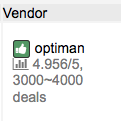}
	\end{center}
	\caption{Reputation of a vendor named `optiman' on Agora.}
\end{figure}

It should be noted that physical items from other continents have a high probability of inspection when going through customs. In the European Union this results in an additional taxes (VAT), which may have implications to the succeeding delivery process.

\section{Delivery of Goods}
The delivery process is the most vulnerable component to the anonymity in the entire purchasing process. Obviously, a delivery address that is linked to the person's identity should never be used. There are two methods that can be used to deliver the physical goods and remain a high level of anonymity. 

\subsection{`Random' drop-off point}
This method consists of using a `random' citizen as man-in-the-middle (MITM). Upon purchase of the goods online, an address other than the buyers personal address is provided to the vendor, with fake commonly occurring name. This address is the address of someone of which the buyer knows it exists and is able to locate. Preferably, this address should be of a person that is home often (typically aged or disabled people) and that is relatively easy reachable from the location of the buyer. Upon purchase, a tracking code is provided by the vendor. This tracking code is used to keep track of the delivery time of the package. As soon as the package is delivered at the MITM address, the buyer confronts the owner of the MITM address pretending to live on the same address, but with a different (but similar, pretending to be a typo) house number, (i.e. 11 instead of 111). The MITM will hand over the package, at which point the buyer returns to its home address, located in a different city.

This method however, has a few weak spots. It is expected that the MITM:
\begin{itemize}
	\item accepts the package, even though not addressed to the name of the MITM.
	\item is home at the moment of delivery. If delivery at the MITM address fails, the entire purchase fails.
	\item does not refrain from opening the package. Depending on the content, this could be fatal for the anonymity delivery, as the MITM may contact law enforcement. 
	\item can not recognise or identify the face of the buyer, during a possible interrogation.
	\item would not ask identification upon handing over the package to the buyer.
\end{itemize}

\subsection{Postal office}
This method consists of using a `random' postal office as man-in-the-middle. This method is more reliable, but hands in a level of anonymity and requires fake documents of identification, which is illegal in almost all countries. As in the previously described method, the address of a postal office together with a fake name, of which the buyer has (fake) identification, would be provided to the vendor. The package is delivered at the postal service and the buyer will retrieve the package, using its fake identification that corresponds to the recipient fake name.

The weak spots in this method are:
\begin{itemize}
	\item the buyer must be in possession of a fake identification.
	\item camera's in the postal office may reveal the buyers real identity.
	\item the package (recipients name) is logged and may be used to de-anonymise the buyer.
	\item multiple fake identities at multiple postal services should be used in order preserve anonymity. 
\end{itemize}

In this method, the postal office may be replaced by a third party post box service that allows physical pick-up of the delivered packages.

\section{Costs of anonymity}
As shown in this paper, anonymity requires many assets that can rapidly result in a  costly operation. In this section, a rough overview is given for the costs of an anonymous operation.

\begin{center}
    \begin{tabular}{ | p{4cm} |  p{4cm} |}
    \hline
    
    \textbf{Component} &  \textbf{Estimated cost} \\ \hline
    Anonymous laptop & \euro{400} \\ \hline
    Air-gapped Raspberry Pi & \euro{50} \\ \hline
    Mobile phone (Nokia 108) & \euro{30} \\ \hline
    SIM card & \euro{10} of which half is credit \\ \hline
    Mixing bitcoins (Bitmixer)& 0.5\% of the total amount plus 0.0005 per forward address.   \\ \hline
    \end{tabular}
\end{center}

This table does not take into account components like fuel for traveling, losses in price fluctuations of bitcoin and foremost, a person's time.

\section{Conclusion}
We have shown that it is possible to preserve a high level of anonymity on the Internet while purchasing physical goods. Good OPSEC practices like compartmentation and encrypted communication is inevitable and essential. Anonymity, as in security, is as strong as its weakest link. We have shown different techniques of obtaining anonymous bitcoins that can be used to purchase virtual services and physical goods. The anonymity is most vulnerable at the delivery process. Finally, we have shown that anonymity comes at a cost that does not make it appealing to execute.

\end{document}